\documentclass[useAMS,usenatbib]{mn2e}
\usepackage{epsfig}
\usepackage{amssymb}

\title[Strong He\,II emission after tidal disruption events] {The production of strong broad He\,II emission after the tidal disruption
of a main-sequence star by a supermassive black hole}
\author[C. M. Gaskell and P. A. Rojas Lobos]{C. Martin Gaskell$^{1,2}$\thanks{E-mail:
martin.gaskell.astro@gmail.com. Present address University of California at Santa Cruz.} and P. Andrea Rojas Lobos$^{1}$\\
\\$^1$Centro de Astrof\'isica de Valpara\'iso y Departamento de F\'isica y Astronom\'ia, Universidad de
Valpara\'iso, \\Av. Gran Breta\~na 1111, Valpara\'iso, Chile.
\\$^2$Department of Astronomy and Astrophysics, University of California,
Santa Cruz, CA 95064\\}

\begin{document}

\date{
Received 2013 September 30. Submitted to Monthly Notices of the Royal Astronomical Society - comments welcome
}

\pagerange{\pageref{firstpage}--\pageref{lastpage}} \pubyear{2013}

\maketitle

\label{firstpage}

\begin{abstract}
The tidal disruption event (TDE) PS1-10jh lacked strong Balmer lines but showed strong, broad, He\,II emission both before maximum light and for at least 8 months thereafter.  Gezari et al.\@ interpreted this as evidence for the disruption of a rare hydrogen-deficient star.  However, Guillochon et al.\@ have argued instead that the disrupted star was a normal main-sequence star and that the strength of the He\,II emission compared with the Balmer lines is a result the emission being similar to the broad-line region (BLR) of an AGN, but lacking the outer, lower-ionization BLR gas.  We show that the profile of He\,II $\lambda$4686 in PS1-10jh is similar to the blueshifted profiles of high-ionization lines in AGNs.  We find an He\,II $\lambda$4686/H$\alpha$ ratio for PS1-10jh of $\sim 3.7$.  We show that both the high-velocity gas of the inner BLR of normal AGNs and the spectra of type II-P supernovae right after shock break out also produce very high He\,II $\lambda$4686/H$\alpha$ ratios.  A high He\,II $\lambda$4686/H$\alpha$ ratio can thus be produced with a solar H/He abundance ratio.  We demonstrate from photoionization modelling that the estimated He\,II $\lambda$4686/H$\alpha$ ratio can be produced with a BLR truncated before the He$^{++}$ Str{\"o}mgren length if the density is $\sim 10^{11}$ cm$^{-3}$. The similarity of the He\,II $\lambda$4686 emission in PS1-10jh to the emission from the inner BLRs of AGNs supports the idea that the emission after a TDE event is similar to that of normal AGNs.
\end{abstract}

\begin{keywords}
tidal disruption events -- galactic nuclei -- emission lines -- blackhole physics -- supernovae.
\end{keywords}

\section{Introduction}

It has long been recognized that stars can be tidally disrupted by supermassive black holes \citep{Hills75,Rees88} and an increasing number of candidate tidal disruption events (TDEs) have been discovered (see \citealt{Komossa12} and \citealt{Gezari12} for recent reviews of observations).  Recently \citet{Gezari+12} obtained the first optical spectrum of a TDE before maximum light.  Contrary to expectations \citep{Bogdanovic+04,Strubbe+Quataert09}, the spectra of this event, PS1-10jh, do {\em not} show strong, broad, irregular, Balmer emission lines, but strong broad He\,II $\lambda$4686 and $\lambda$3203 emission instead.  \citet{Gezari+12} therefore proposed that the PS1-10jh event was due to the disruption of a helium star (see \citealt{Bogdanovic+13} for further discussion of a possible scenario).  However, \citet{Guillochon+13} have argued that this is highly unlikely because helium-rich stars are rare in the universe.  They argue instead that a TDE produces a temporary accretion disc analagous to the longer-lived accretion disc in a ``normal'' thermal AGN (see \citealt{Antonucci12} for a discussion of thermal versus non-thermal AGNs) and that the broad He\,II emission arises from a temporary broad-line region (BLR) above the accretion disc as in a normal thermal AGN (see \citealt{Gaskell09} for a review of properties of the broad-line region).  \citet{Gaskell+07}(GKN) argue that the BLR in normal AGNs is self-shielding and that, as a consequence, the ionization is highly radially stratified.  In the GKN model the He\,II emission is effectively entirely produced in an inner region which is separate from the outer region in which the Balmer line emission predominantly arises.  This structure is strongly supported by observations of the BLR (see review of \citealt{Gaskell09}).  \citet{Guillochon+13} have proposed that the unusually large He\,II $\lambda$4686/H$\alpha$ ratio in PS1-10jh is a consequence of a temporary BLR above the accretion disc of the TDE only having the inner, high-ionization part of the BLR and not the outer, lower-ionization region.

In this paper we compare the profile of He\,II $\lambda$4686 with the profile of high-ionization broad lines in AGNs and we estimate the strength of H$\alpha$ and broad other lines.  We show that He\,II $\lambda$4686/H$\alpha$ ratio for the high-ionization inner part of the BLR in AGNs is very large as is also the ratio in the spectra of type-II supernovae (SNe) right after shock break out.  We then use the photoionization code CLOUDY to model the He\,II $\lambda$4686/H$\alpha$ ratio for matter-bounded photoionized clouds and demonstrate how the high He\,II $\lambda$4686/H$\alpha$ ratio observed in PS1-10jh, the inner BLR of AGNs and in the earliest time spectra of SNe II-P can readily be produced with a solar He/H ratio.

\section{The pre-maximum-light spectrum of PS1-10jh}.

The TDE PS1-10jh has been the best studied TDE to date.  \citet{Gezari+12} obtained a well-sampled light curve starting over 50 days before maximum light (see their Figure 2) and obtained optical spectroscopy with the MMT 22 days before maximum light.  In their Figure 1 they show the pre-maximum-light spectrum with and without the host galaxy spectrum subtracted (see their Supplementary Information for details).  We fit a continuum to line-free spectral regions of this spectrum using a fourth-order polynomial. The resulting continuum-subtracted spectrum is shown in Figure 1.  To facilitate comparisons with models we have smoothed the spectrum.  For rest-frame wavelengths shortwards of 5140\AA ~we smoothed the spectrum with a boxcar of half-width 37\AA~(corresponding to $\sim \pm \,1700$ km s$^{-1}$ at He\,II $\lambda$4686) .  The signal-to-noise ratio is substantially lower in the red end of the spectrum so we smoothed the spectrum longwards of 5140\AA ~by 120\AA ~($\sim \pm \,5400$ km s$^{-1}$ at H$\alpha$).

\begin{figure}
 \centering \includegraphics[width=8.5cm]{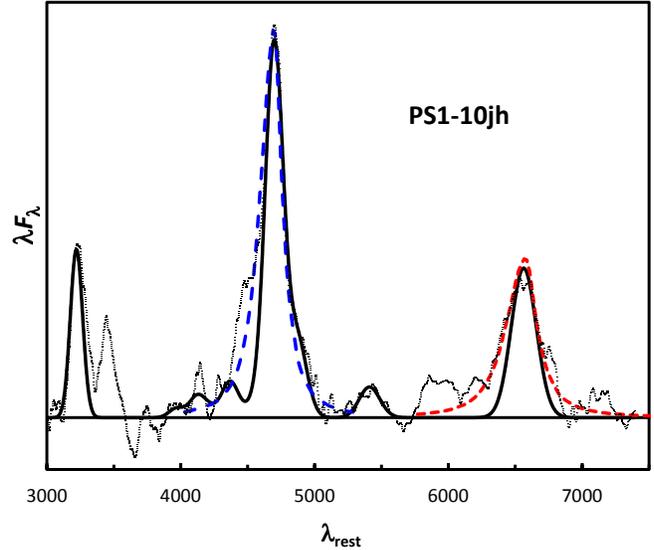}
 \caption{The thin dotted line is the MMT spectrum of PS1-10jh taken 22 days before maximum light after the host-galaxy subtraction of \citet{Gezari+12} and after our additional subtraction of the thermal continuum and smoothing (see text).  The horizontal line indicates the zero level.  The solid black curve is a synthetic He\,II + Balmer lines fit using a single Gaussian.  The width of the Gaussian has been fit to core of the He\,II $\lambda$4686 profile. The intensities of He\,II $\lambda$4686 and H$\alpha$ have been fit independently, but the weaker He\,II and hydrogen Balmer lines have been assumed to have their case B values.  The long-dashed blue curve is the blueshifted AGN profile from Figure 7 of \citet{Gaskell+Goosmann13} with the intensity and width fit to He\,II $\lambda$4686 (this does not included the additional weaker blended lines).  The short-dashed red curve is the same profile shifted to the wavelength of H$\alpha$.}
\end{figure}

\subsection{Helium and hydrogen line intensities}

In addition to the broad $\lambda$4686 and $\lambda$3203 He\,II  Fowler series lines previously identified by \citet{Gezari+12} we can now also clearly see a similarly broad line at $\sim \lambda$6560.  Alternate members of the Pickering series of He\,II coincide with Balmer lines of hydrogen so some of the emission at $\lambda$6560 is He\,II.  Because they come from high energy levels, the He\,II line ratios will be close to Case B values (see, for example, \citealt{Osterbrock+Ferland06}) and we estimate that the Pickering line contributes 20\% to the $\lambda$6560 blend.  This gives an He\,II $\lambda$4686/H$\alpha$ ratio of 3.7 with an uncertainty of around 25\%.  We show the expected Case B intensities of the other Fowler, Pickering, and Balmer lines in Figure 1. For purposes of illustration we have assumed that the profiles can all be represented by Gaussians with FWHMs of 10,000 kms$^{-1}$.  It can be see that the intensities of He\,II $\lambda$3203 and the possible He\,II $\lambda$5411 line are correctly predicted.  H$\beta$ and the corresponding Pickering line are a cause of the inflexion in the red wing of He\,II $\lambda$4686 and addition Balmer and Pickering lines contribute to the extended blue wing.

\subsection{He\,I $\lambda$5876 and Ne\,V $\lambda$3426?}

Because of the noise in the spectrum, and the dangers of spurious features due to inevitable slight mismatches in the galaxy template subtraction, weak features in the spectrum should not be over interpreted.  The broad emission around $\lambda$5880 does not correspond to He\,II line so, if it is real, we suggest it could be He\,I $\lambda$5876.  The signal-to-noise ratio is worst at the extremes of the spectrum.  The apparent feature at $\lambda$7100 has no obvious identification and we assume it is just noise.  At the blue end of the spectrum there is also greater sensitivity to mismatches in the galaxy subtraction.  An apparent broad emission at $\sim \lambda$3430 agrees in wavelength of Ne\,V $\lambda$3426, but we consider this identification unlikely since there are no other obvious forbidden lines in the spectrum with similar ionizations or critical densities.  We  therefore consider this feature to be spurious.  Note that there is an apparent absorption feature nearby at $\lambda$3650 of similar strength which also has no obvious identification and is probably noise.

\subsection{The He\,II emission line profile}

The Gaussian used in Figure 1 is clearly not a good fit to the blue side of He\,II $\lambda$4686.  Other Balmer and He\,II lines are not at the right wavelength to explain the profile.  He\,I $\lambda$4471 has the right wavelength but it is much weaker than He\,I $\lambda$5876 so it cannot explain the blue wing.  However, the high-ionization lines of normal AGNs commonly show strongly blue asymmetric line profiles \citep{Gaskell82}.  This has been proposed to be a result of the combination outflow of the high-ionization BLR and obscuration by the accretion disc suppressing the red side of the line \citep{Gaskell82}.  Although this proposal has become popular, velocity-resolved reverberation mapping of high-ionization lines fails to show the expected outflow signature \citep{Gaskell88,Koratkar+Gaskell89,Crenshaw+Blackwell90,Koratkar+Gaskell91,Ulrich+Horne96,Kollatschny03}.  In fact, velocity-resolved reverberation mapping favours a net {\em inflow} of the BLR (see \citealt{Gaskell10} and \citealt{Gaskell+Goosmann13}).  Scattering off inflowing material will naturally produce a blueshifted profile \citet{Gaskell09,Gaskell+Goosmann13} and the predicted profiles are in good agreement with observed profiles of high-ionization BLR lines.  In Figure 1 we show a theoretical scattering profile of \citet{Gaskell+Goosmann13} compared with the observed profiles of He\,II$\lambda$4686 and H$\alpha$.  As can be seen, the agreement is satisfactory within the observational uncertainties.   The FWHM of this blueshifted profile is $\sim$ 11,000 km s$^{-1}$ which is similar to what is observed in AGNs (see Figure 6 of \citealt{Gaskell+Goosmann13}).   We therefore suggest that the profile of He\,II $\lambda$4686 in PS1-10jh, like the profiles of high-ionization BLR lines in AGNs, is due to scattering off inflowing material.

\section{The He\,II $\lambda$4686/H$\alpha$ ratio in normal AGNs}

\citet{Guillochon+13} argue that the strong He\,II emission seen in the pre-maximum spectrum of PS1-10jh comes from a region similar to the inner BLR of AGNs, and we have argued above that the line profile is consistent with what is seen in AGNs.  We would therefore expect the He\,II $\lambda$4686/H$\alpha$ ratio to be similar to that seen in AGNs.  The measurements of \citet{Osterbrock77}, give He\,II $\lambda$4686/H$\alpha$ $\sim 0.12$ on average for normal AGNs while for the pre-maximum spectrum of PS1-10jh we obtain $3.7 \pm 0.9$.  A key question is: is the observed ratio in PS1-10jh consistent with solar abundances given this difference of one and a half orders of magnitude from the mean ratio for AGNs?

As has been normal in AGN studies, the measurements of \citet{Osterbrock77} are for {\em integrated} line profiles. Furthermore, his measurements assumed that He\,II $\lambda$4686 had the same profile as H$\beta$.  The ratio of 0.12 thus refers to an average weighted towards the regions producing H$\beta$.  However, the high-ionization broad lines in AGNs do {\em not} have the same profiles as the Balmer lines and low ionization lines.  The high-ionization lines are substantially broader \citep{Osterbrock+Shuder82,Shuder82,Mathews+Wampler85} and the widths of lines progressively increases with ionization (see \citealt{Gaskell09}).  An obvious point that has not received much attention is that because of the different widths, {\em the line ratios are substantially different in the wings}.   For example, in the profile comparisons of \citet{Shuder82} and \citet{Crenshaw86} the He\,I $\lambda$5876/H$\beta$ ratio is 3 -- 5 times higher in the wings.  Most photoionization modelling of the BLR has focussed on integrated line profiles and there has been only a little work modelling line ratios as a function of velocity (e.g., \citealt{Snedden+Gaskell07}).  Unlike the relatively unblended He\,I $\lambda$5876 line, the profile of He\,II $\lambda$4686 has not been studied much. However, \citet{Osterbrock+Shuder82} present profiles of He\,II $\lambda$4686 three AGNs.  In all three cases, He\,II $\lambda$4686 shows a substantial blueshift. In Figure 2 we show the He\,II $\lambda$4686/H$\alpha$ ratio as a function of velocity for Mrk 335, which has the best profiles in \citet{Osterbrock+Shuder82}.  It can be seen the the ratio is substantially higher in the wings than in the line center.   As discussed by \citet{Shuder82}, line profile ratios in the wings are sensitive to the placement of the continuum level. It is probable that the dip in the ratio at high positive velocities is a consequence of the choice of continuum level (see Figure 2 of \citealt{Shuder82}).  Although much noisier, the other two He\,II profiles in \citet{Osterbrock+Shuder82} are consistent with the increase in He\,II $\lambda$4686/H$\alpha$ at high velocities that we show in Figure 2.

\begin{figure}
 \centering \includegraphics[width=8.5cm]{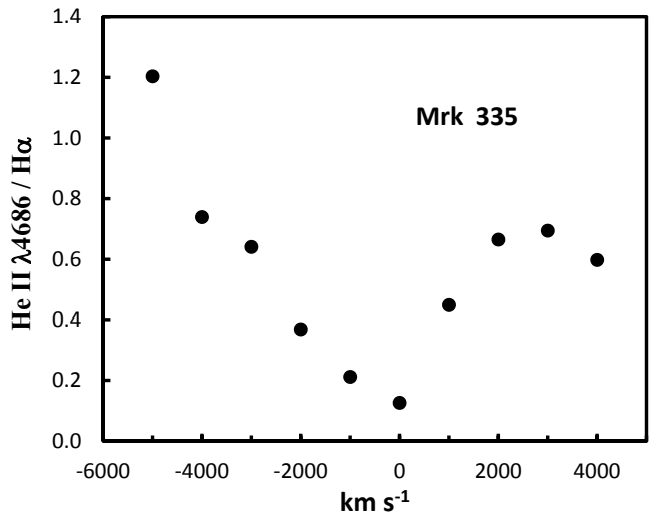}
 \caption{The He\,II $\lambda$4686/H$\alpha$ ratio as a function of velocity for the broad-line region of Mrk 335.}
\end{figure}

From Figure 2 we can see that the He\,II $\lambda$4686/H$\alpha$ ratio in the high-velocity blue wing of Mrk 335 begins to approach the ratio we find for PS1-10jh.  This shows that {\em strong helium emission does not require depletion of hydrogen}.

\section{The earliest time optical spectra of type II-P supernovae}

Type II supernovae (SNe II), by definition, show strong hydrogen emission lines in their spectra.  However, spectra of SNe II-P taken right after shock breakout (within a few days) lack strong hydrogen emission lines.  The strongest feature is broad emission around $\lambda$4600 \citep{Lewis+94,Leonard+02,Quimby+07}.  This feature, which is particularly well shown in Figure 6 of \citet{Leonard+02}, is initially much more prominent than H$\alpha$ (see the spectra of day 3 of SN 1993J in Figure 4 of \citealt{Lewis+94}, day 1 of SN 1999gi in Figure 5 of \citealt{Leonard+02}, and day 2 of SN 2006bp in Figure 6 and 8 of \citealt{Quimby+07}).   \citet{Baron+00} failed to find a convincing identification for the $\lambda$4600 feature.  \citet{Leonard+02} associated the feature with a highly blueshifted component of a complex P Cygni profile of H$\beta$ but, as \citet{Quimby+07} pointed out, a problem with this interpretation is the lack of a corresponding high-velocity H$\alpha$ feature.  \citet{Quimby+07} demonstrated instead that the $\lambda$4600 feature is blueshifted He\,II $\lambda$4686.  They showed that the rapid decline in the strength of He\,II $\lambda$4686 in the first few days after shock breakout is coupled with an increase in strength of He\,I $\lambda$5876 as expected as the ionization declines (see their Figures 6 and 7).  

Because the spectra of the above mentioned SNe II-P all soon show strong hydrogen emission, the days right after shock breakout in SNe II-P thus present another example of how a high He\,II $\lambda$4686/H$\alpha$ ratio similar to that observed in PS1-10jh can be produced without having to invoke depletion of hydrogen.

\section{Modeling He\,II $\lambda$4686 emission}

The GKN model of the BLR of normal AGNs differs from previous models of the BLR (such as the locally optimized cloud model of \citealt{Baldwin+95}) in that the BLR is self-shielding so that the low-ionization gas sees the ionizing continuum through the inner, high-ionization gas (see Figures 1 -- 3 of \citealt{Gaskell09} for an explanation).  This produces a high degree of radial ionization stratification of the BLR, as is found through reverberation mapping (see \citealt{Gaskell09}).  \citet{Guillochon+13} propose that PS1-10jh only has the inner, high-ionization part of the BLR.  We used version 13.01 of the photoionization code CLOUDY \citep{Ferland+98,Ferland+13} to calculate theoretical He\,II $\lambda$4686/H$\alpha$ ratios for the GKN model to quantitatively investigate the proposal of Guillochon et al.  For the ionizing continuum we simply adopted the standard AGN continuum of \citet{Mathews+Ferland87}.\footnote{For our purposes, so long as there are photons with energies greater than 4 Rydbergs to ionize He$^+$, the precise shape of the spectrum is unimportant since there is a degeneracy between the ionization parameter and the spectral energy distribution.}  We ran constant density models because there would have been insufficient time to achieve pressure equilibrium in the stellar debris before the epoch of the first spectrum 22 days before maximum light.

Traditional, photoionization modelling of AGN spectra has assumed that BLR clouds have a sufficiently large column density that the size of a cloud is substantially greater than the hydrogen Str{\"o}mgren length in order to produce the wide range of ionization seen (see, for example, the review of \citealt{Davidson+Netzer79}).  In contrast to this, in the GKN model the ensemble of clouds in the self-shielding BLR is treated as an expanded single cloud (see Figure 4 of \citealt{Gaskell09}).  In Figure 3 we show the He\,II $\lambda$4686/H$\alpha$ ratio for such a cloud truncated at various Str{\"o}mgren lengths.   Beyond one hydrogen Str{\"o}mgren length the cloud is effectively radiation-bounded and the He\,II $\lambda$4686/H$\alpha$ ratio declines very slowly since the Balmer line emission is much less beyond that point.  Such a cloud corresponds to the situation traditionally used to model integrated BLR line intensities and the calculated He\,II $\lambda$4686/H$\alpha$ ratio is close to (but slightly less than) the observed integrated ratio.  However, He\,II $\lambda$4686 primarily comes from the front of the cloud where helium is doubly ionized. The He\,II $\lambda$4686/H$\alpha$ ratio will be a maximum if the cloud is truncated at the end of the He$^{++}$ Str{\"o}mgren zone.  The ratio will then decline from that point to the hydrogen Str{\"o}mgren length because the H$\alpha$ emission is continuing to increase while there is little additional He\,II $\lambda$4686 emission.  This can be seen in Figure 3.

\begin{figure}
 \centering \includegraphics[width=8.5cm]{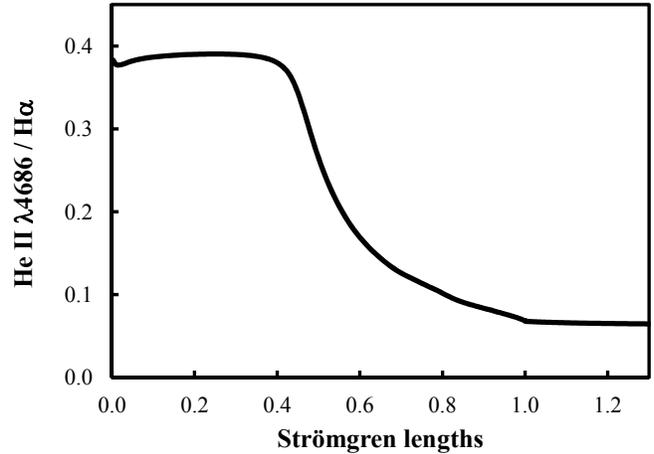}
 \caption{The cumulative ratio of He\,II$\lambda$4686/H$\alpha$ as a function of the size of a matter-bounded cloud in terms of the hydrogen Str{\"o}mgren length for the low-density limit ($n_H < 10^{8}$ cm$^{-3}$) when the cloud is ionized by the AGN continuum of \citet{Mathews+Ferland87}.}
\end{figure}

Our calculations show (see Figure 3) that truncating the BLR in the He$^{++}$ Str{\"o}mgren zone does indeed produce a significantly higher He\,II $\lambda$4686/H$\alpha$ ratio ($\approx 0.4$), but this is still much lower than the observed value in the innermost BLR (see Figure 2) or in the earliest spectra of SNe II-P and in PS1-10jh.  Truncating the BLR at the end of the He$^{++}$ zone is not in itself sufficient to produce He\,II $\lambda$4686/H$\alpha > 1$.  We therefore investigated changing the physical conditions.  Since we are only considering the He$^{++}$ Str{\"o}mgren region where, by definition, He is almost entirely He$^{++}$ and H almost entirely H$^+$, the He\,II $\lambda$4686/H$\alpha$ ratio depends only on the relative He/H abundance, the ratio of the emissivities, and the probabilities of escape of the photons. We therefore do not expect the He\,II $\lambda$4686/H$\alpha$ ratio to depend on the ionization parameter (the ratio of ionizing photons to gas density) or the spectral shape and we confirmed this by running models with $-2 < \log U < 2$. However, we found a substantial variation in the ratio as a function of density (see Figure 4).  This is because of the increasing thermalization of the lines at high densities as the spectrum tends towards the black body limit.  At a given physical depth the optical depths in the Balmer lines is more than an order of magnitude higher than for the Fowler lines because (a) the abundance of hydrogen is ten times higher and (b) the relative populations of the excited states of hydrogen are higher compared with He$^+$ because of the energy levels of hydrogen for similar principal quantum numbers are four times lower.  As the density rises, thermalization will first reduce the relative strength of the Balmer lines, causing the He\,II $\lambda$4686/H$\alpha$ ratio to increase.  Then, at higher densities still, there will be a reduction of the strength of the He\,II emission which will lower the He\,II $\lambda$4686/H$\alpha$ ratio again.  These effects can be seen in Figure 4 and it can be seen that a density of $\sim 10^{11}$ cm$^{-3}$ produces the observed He\,II $\lambda$4686/H$\alpha$ ratio.

\begin{figure}
 \centering \includegraphics[width=8.5cm]{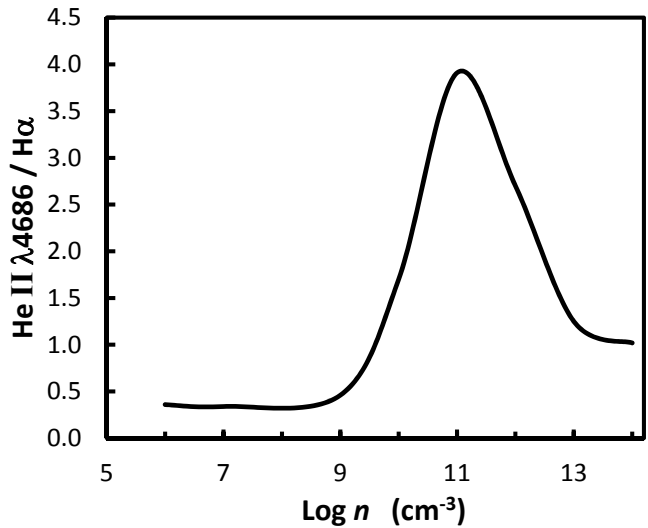}
 \caption{The maximum ratio of He\,II$\lambda$4686/H$\alpha$ for matter-bounded clouds as a function of density, $n_H$.}
\end{figure}

\section{Conclusions}

We have derived the He\,II $\lambda$4686/H$\alpha$ ratio from the pre-maximum light spectra of the candidate tidal disruption event PS1-10jh and shown that the He\,II $\lambda$4686 profile is similar to the blueshifted high-ionization BLR lines of normal AGNs. We have pointed out that both the inner, high-velocity BLR of normal AGNs, and the ejecta of type II-P supernovae right after shock breakout also show a high He\,II $\lambda$4686/H$\alpha$ ratio and we are able to reproduce this ratio with photoionization models with solar abundances so long as the density is $\sim 10^{11}$ cm$^{-3}$. This provides strong support for the suggestion of \citet{Guillochon+13} that the star disrupted in PS1-10jh was a common main sequence star rather than a very rare helium star.  Finally, we note that the similarity of the He\,II $\lambda$4686 emission in PS1-10jh to the emission from the inner BLRs of normal AGNs supports the idea that the emission after a TDE event is a temporary version of the emission in normal thermal AGNs.

\section*{Acknowledgments}

We are grateful to Enrico Ramirez-Ruiz and James Guillochon for stimulating our interest in the interpretation of the spectrum of PS1-10jh and for sharing their results in advance of publication.  This research has been supported in part by FONDECYT of Chile through project N{\degr} 1120957 and the GEMINI-CONICYT Fund of Chile through project N{\degr}32070017.


\end{document}